\documentclass[runningheads,orivec]{llncs}
\usepackage[T1]{fontenc}
\usepackage{amsmath}
\usepackage{amsfonts}
\usepackage{graphicx}
\usepackage{soul}
\usepackage{multirow}
\usepackage{multicol}
\usepackage{booktabs}
\usepackage{colortbl}
\usepackage{xspace}
\usepackage[hidelinks]{hyperref}

\usepackage{cleveref}
\usepackage[%
  square,        
  comma,         
  numbers,       
  sort&compress,
]{natbib}

\Crefname{equation}{Eq.}{Eqs.}
\Crefname{figure}{Fig.}{Figs.}
\Crefname{table}{Tab.}{Tabs.} 
\Crefname{appendix}{App.}{Apps.} 
\crefformat{section}{\S#2#1#3}
\crefformat{subsection}{\S#2#1#3}

\begin{document}
\title{Measuring Individual User Fairness with User Similarity and Effectiveness Disparity}
\titlerunning{Measuring Individual User Fairness}
\author{Theresia Veronika Rampisela\inst{1}\orcidID{0000-0003-1233-7690} 
\and Maria Maistro\inst{1}\orcidID{0000-0002-7001-4817} 
\and Tuukka Ruotsalo\inst{1,2}\orcidID{0000-0002-2203-4928} 
\and Christina Lioma\inst{1}\orcidID{0000-0003-2600-2701}}
\authorrunning{Rampisela et al.}

\institute{University of Copenhagen, Copenhagen, Denmark \email{\{thra,mm,tr,c.lioma\}@di.ku.dk}
\and
LUT University, Lappeenranta, Finland\\
}

\maketitle              
\begin{abstract}
Individual user fairness is commonly understood as treating similar users similarly. In Recommender Systems (RSs), several evaluation measures exist for quantifying individual user fairness. These measures evaluate fairness via either: (i) the disparity in RS effectiveness scores regardless of user similarity, or (ii) the disparity in items recommended to similar users regardless of item relevance. Both disparity in recommendation effectiveness and user similarity are very important in fairness, yet no existing individual user fairness measure simultaneously accounts for both. 
In brief, current user fairness evaluation measures implement a largely incomplete definition of fairness. 
To fill this gap, we present Pairwise User unFairness (PUF), a novel evaluation measure of individual user fairness that considers both effectiveness disparity and user similarity. 
PUF is the only measure that can express this important distinction. We empirically validate that PUF does this consistently across 4 datasets and 7 rankers, and robustly when varying user similarity or effectiveness. 
In contrast, all other measures are either almost 
insensitive to effectiveness disparity or completely insensitive to user similarity. We contribute the first RS evaluation measure to reliably capture both user similarity and effectiveness in individual user fairness. Our code: \href{https://github.com/theresiavr/PUF-individual-user-fairness-recsys}{github.com/theresiavr/PUF-individual-user-fairness-recsys}. 

\end{abstract}
\newcommand{\norm}[1]{\left\lVert#1\right\rVert}

\section{Introduction}
\label{s:intro}

Recommender System (RS) evaluation has always relied heavily on effectiveness as it directly affects user utility and satisfaction. Existing work on RS fairness evaluation often uses measures that depend on effectiveness scores, especially when it comes to individual user fairness. \emph{Individual fairness} is traditionally defined as treating similar individuals similarly \cite{Dwork2012FairnessAwareness}. 
We focus on attribute-free fairness for individual users, where we assume no user attribute is available, other than the user identifiers and interactions \cite{Li2024ExplainingPerspective,Zeng2024FairDemographics}. Sensitive attributes (e.g., race, age) are frequently unavailable due to privacy or data incompleteness. 

To measure individual user fairness in RSs, the disparity in recommendation effectiveness across users is often used as a proxy of how similarly the recommendation algorithm treats the users. 
Recommendations across users are deemed fair if similar users have similar effectiveness scores. Yet, no existing individual user fairness measure considers both effectiveness and user similarity (see below):

    \begin{center}           \resizebox{0.7\columnwidth}{!}{
    \begin{tabular}{llcc}
         \toprule
         Measure & Reference & Effectiveness & User Similarity \\
         \midrule
         Standard deviation & \cite{Patro2020FairRec:Platforms,Biswas2021TowardPlatforms,Wu2021TFROM:Providers,Rastegarpanah2019FightingSystems,Li2024ExplainingPerspective,Xu2024PromotingFairness}   & $\checkmark$ & -\\
         Gini index & \cite{Fu2020Fairness-AwareGraphs,Leonhardt2018UserSystems} & $\checkmark$ & - \\
         Envy-based &\cite{Patro2020FairRec:Platforms,Biswas2021TowardPlatforms,Do2022OnlineSystems}& $\checkmark$ & -\\
         UF & \cite{Wu2023EquippingEmbedding} &- & $\checkmark$  \\
         \midrule
         PUF & ours & $\checkmark$ & $\checkmark$ \\
         \bottomrule
    \end{tabular}}
    \end{center}

User similarity and recommendation effectiveness are both important, as the former affects our expectation of how close effectiveness scores should be. 
E.g., given two users whose past interactions are highly similar, the recommendations they receive are deemed fair if their effectiveness is similar, and vice versa. This is because fairness means similar users should get similar treatment. However, two dissimilar users cannot expect to get similar treatment, as their past interactions may differ, e.g., in terms of amount, frequency, or item type.

Current individual user fairness measures can be grouped into those that consider: (i) only effectiveness disparity; and (ii) only user similarity and recommendation similarity. For (i), different effectiveness for users is not always unfair. Some users may have only a few past interactions, and others may have very specific tastes for whom only a few items are relevant. Fairness should consider that these users are different and cannot be compared to users who, e.g., consume only popular items.
For (ii), recommending similar items to similar users may not be fair if one user is satisfied with their recommendation, but another is not. In this case, the recommendation is not really fair as its effectiveness differs.

To counter the above limitations, we propose a novel evaluation measure for individual user fairness in RS: Pairwise User unFairness (PUF). PUF quantifies individual user fairness based on the disparity in recommendation relevance between user pairs, weighted by the similarity of user pairs. As such, PUF is aligned with the definition of individual fairness and also accounts for recommendation effectiveness, and thus, user utility. We show that compared to existing measures, PUF has higher sensitivity to changes in effectiveness scores and user similarity distribution.  
Overall, we contribute a new evaluation measure for individual user fairness, which considers both user similarity and disparity in recommendation effectiveness, and which does not have the same limitations as existing measures.

\section{Individual user fairness}
\label{s:individual-user-fair}
We define individual user fairness as per \cite{Dwork2012FairnessAwareness}: let $u$ and $u'$ denote two users; $L_u$ and $L_{u'}$ be the recommendation lists of these two users; and $M(\cdot)$ be a function mapping a recommendation list to a score, e.g., its effectiveness score. Any two users whose profile distance is $d(u,u')$ should receive recommendations such that recommendation effectiveness satisfies $D(M(L_u), M(L_{u'})) \leq d(u,u')$, where $D$ is a distance measure. 
In other words, fairness is achieved when the difference in the users' recommendation effectiveness is at most $d(u,u')$. This definition agrees with the definitions of RS individual user fairness in \cite{Zehlike2022FairnessSystems,Smith2023ScopingPerspective}. 

Next, we present all evaluation measures of attribute-free individual user fairness in RSs that have been published up to January 2025 \cite{Wang2022,Amigo2023ASystems,Smith2023ScopingPerspective,LiYunqi2023FairnessApplications,Zhao2024FairnessSurvey,Wu2023FairnessStrategies,Aalam2022EvaluationReview,Zehlike2022FairnessSystems,Pitoura2022FairnessOverview}; we include their equations in an online appendix together with the code repository.  
All of them quantify unfairness, and the lower their score, the fairer (denoted by $\downarrow$). 
All these measures, except the distance-based measure, quantify effectiveness disparity but ignore user similarity. 
While the distance-based measure considers user similarity, it is detached from effectiveness. Hence, no existing individual user fairness measure in RS considers both user similarity and effectiveness disparity. 

\noindent\textbf{Standard deviation (SD)}. 
In RSs, $\downarrow$SD and variance are often used to quantify individual user fairness. An RS is fair if it provides equal prediction accuracy to all users, and fairness is evaluated via the variance of the mean squared error (MSE) of user rating prediction \cite{Rastegarpanah2019FightingSystems,Li2024ExplainingPerspective}. Other works measure fairness from user recommendation lists, e.g., via the variance of NDCG scores across users \cite{Wu2021TFROM:Providers} or the SD of user utility \cite{Patro2020FairRec:Platforms,Biswas2021TowardPlatforms,Xu2024PromotingFairness}. 

\noindent\textbf{Gini Index (Gini)}. $\downarrow$Gini is a well-known inequality measure that quantifies the extent to which a distribution deviates from a perfectly equal distribution. It has been used to measure individual user fairness from different distributions, e.g., P@$k$ (Gini-P), NDCG (Gini-NDCG), or the utility score per user \cite{Fu2020Fairness-AwareGraphs,Leonhardt2018UserSystems}.

\noindent\textbf{Envy-based measures.} Envy is defined as the extra utility a user $u$ would have received if they were given the recommendation list of user $u'$, $L_{u'}$ \cite{Patro2020FairRec:Platforms,Biswas2021TowardPlatforms}. 
Three fairness evaluation measures are based on this concept, where envy is aggregated differently: Mean Envy ($\downarrow$ME) \cite{Patro2020FairRec:Platforms,Biswas2021TowardPlatforms}, Mean Max Envy ($\downarrow$MME) \cite{Do2022OnlineSystems}, and Proportion of $\epsilon$-Envious User ($\downarrow$PEU) \cite{Do2022OnlineSystems}. 

\noindent\textbf{Distance-based measures}.
In \cite{Wu2023EquippingEmbedding}, fairness for individual users is defined as any two similar users $u,u'$ receiving similar recommendations. Recommendation disparity is then measured with UnFairness score ($\downarrow$UF). UF uses both the user similarity and the pairwise distance between the representation (e.g., embeddings) of items in the recommendation list of users $u$ and $u'$. User similarity is modelled by a weighted sum of Jaccard ($sim_{Jacc}$) of the user's set of past interactions and JS-div between item feature (e.g., genre) distributions in the interactions of user $u$ and $u'$. UF does not consider recommendation effectiveness. 

\section{Pairwise User unFairness (PUF)}
\label{s:puf-measure}

We present our evaluation measure of individual user fairness, Pairwise User unFairness ($\downarrow$PUF). PUF has two components: similarity between users and disparity in recommendation effectiveness, which we describe next.

\noindent \textbf{(1) Measuring similarity.} Measuring similarity between users is an inherently hard problem and there is no single ground truth of what makes two users similar \cite{Buyl2024InherentFairness}. When there is no user attribute, as in our case, user profiles can be modelled based on their historical interactions \cite{Herlocker1999AnFiltering,Wu2023EquippingEmbedding,Li2024ExplainingPerspective}. User similarity is then computed pairwise based on the user representation (e.g., click/rating matrix, user embedding), for example, with cosine similarity \cite{Reisz2024QuantifyingPrediction} or Jaccard \cite{Wu2023EquippingEmbedding}.

\noindent \textbf{(2) Measuring disparity.} 
PUF quantifies individual user fairness based on disparity in recommendation effectiveness, considering user similarity. 
PUF measures the mean pairwise difference in the effectiveness score per user, weighted by the user pair similarity. 
Based on (1) and (2), we define PUF as follows:
\begin{equation}
    \text{PUF} = \frac{2}{m(m-1)}
    \sum_{u\in U}\sum_{u'\in U\setminus\{u\}} 
    sim(u,u') \times  |S(u) - S(u')|
\end{equation}
\noindent where $U$ is the set of all users ($|U|=m$, where $m\geq2$), 
$sim(u,u') \in [0,1]$ is the similarity of users $u$ and $u'$, $S$ is an effectiveness measure, e.g., P@$k$. 
$S$ must range or scaled to be in $[0,1]$, so PUF also ranges in $[0,1]$. PUF can be used with any similarity/effectiveness measure that fulfils the range requirement. 

\noindent \textbf{How PUF differs from UF.} 
While both PUF and UF \cite{Wu2023EquippingEmbedding} consider user similarity, PUF considers the difference in the recommendation relevance, rather than only the disparity based on the representation of the recommended items as in UF. Moreover, recommending different sets of items to two similar users may be considered unfair by UF even if both users like their recommendations, but it is fair based on PUF. 
In theory, users with similar tastes in the past are likely to have a similar preference in the future \cite{Resnick1994GroupLens:Netnews}. Yet, the recommendation problem is challenging as even two highly similar users may not equally like their recommended items, if they are given identical items. This may be due to, for example, incomplete historical data that is unrepresentative of user taste, diverging user preference, or limited ground truth data. These limitations necessitate a look into the disparity of the recommendation relevance, which is what our measure quantifies. 

Overall, our measure, PUF, aligns with the definition of individual user fairness and quantifies fairness through the disparity in recommendation effectiveness, which is more meaningful than the disparity in the representation of recommended items, as effectiveness relates more to user utility.

\section{Empirical analysis}
\label{s:experiment}

We compare PUF to existing effectiveness and individual user fairness measures. 
\subsection{Experimental setup}
\label{ss:setup}
\noindent \textbf{Datasets}. We use four real-world datasets from three domains: music (Lastfm \cite{Cantador20112nd2011}), video (QK-video \cite{Yuan2022Tenrec:Systems}), and movie (ML-10M and ML-20M \cite{Harper2015TheContext}). 
We obtain Lastfm and ML-* from \cite{Zhao2021RecBole:Algorithms}, and QK-video from \cite{Yuan2022Tenrec:Systems}. We use the `sharing' interactions in QK-video. For all datasets, we remove duplicate interactions and keep the most recent. We remove users and items with $<5$ interactions. For ML-* we map ratings $\geq3$ to 1, and discard the rest. The threshold 3 is chosen as the ratings range between $[0.5, 5]$. Lastfm and QK-video have unary ratings, so no mapping is required. We split the preprocessed datasets into train/val/test with a ratio of 6:2:2. 
The ML-* datasets are temporally split, while Lastfm and QK-video are randomly split as they have no timestamps. We split datasets globally (not user-wise) to avoid data leakage \cite{Meng2020ExploringModels}. After splitting, users with $<5$ interactions in the train set are removed from all splits to ensure that each user has adequate training data. The final preprocessed dataset statistics are in \Cref{tab:stats}. 

\noindent \textbf{Recommenders}. We use 7 well-known collaborative filtering recommenders: user- and item-based $K$-nearest neighbour (U-KNN \cite{Resnick1994GroupLens:Netnews} and I-KNN \cite{Deshpande2004Item-basedAlgorithms}), Bayesian Personalised Ranking (BPR \cite{RendleBPR:Feedback}), Variational Autoencoder with multinomial likelihood (MVAE \cite{Liang2018VariationalFiltering}), Neural Graph Collaborative Filtering (NGCF \cite{Wang2019NeuralFiltering}), Neural Matrix Factorisation (NMF \cite{He2017NeuralFiltering}), and Neighbourhood-enriched Contrastive Learning (NCL \cite{Lin2022ImprovingLearning}). All models except U- and I-KNN are trained for 300 epochs with early stopping. We tune hyperparameters with grid search. The configuration with the best NDCG@10 during validation is the final model. Implementation, training, and tuning are done with RecBole \cite{Zhao2021RecBole:Algorithms}.

\noindent \textbf{Evaluation measures}. We measure recommendation effectiveness (\textsc{Eff}) with Hit Rate (HR), MRR, Precision (P or Prec), Recall (R), MAP, and NDCG \cite{Jarvelin2002CumulatedTechniques}. Individual user fairness (\textsc{Fair}) is evaluated with our PUF measure (\Cref{s:puf-measure}) and all existing measures (\Cref{s:individual-user-fair}): standard deviation (SD) of the P@$k$ (SD-P) and NDCG@$k$ (SD-NDCG) scores across all users; Gini Index of P@$k$ (Gini-P) and NDCG@$k$ (Gini-NDCG); envy-based measures (ME, MME, and PEU); and distance-based measure, UF. 

We evaluate all runs at $k=10$. For PEU, we set $\epsilon=0.05$ 
\cite{Do2022OnlineSystems}. For UF, the number of all user pairs is used as the log base, so that UF $\in[0,1]$. As we use several models with diverse item representations, 
for a fair comparison between all models, we represent items with the one-hot encoding of the historical interactions. We test four variants of PUF (two user similarity measures $\times$ two effectiveness measures). We use cosine ($sim_{cos}$) and Jaccard ($sim_{Jacc}$), for similarity measures as both are used in UF.\footnote{For UF and PUF, we also vary the weights of user interaction and item features in computing user similarity. As we find similar conclusions, we report only interaction-based similarity; the rest is in the online appendix.}  
In addition, $sim_{cos}$ is used in our U-KNN model, as well as to analyse user similarity \cite{Reisz2024QuantifyingPrediction}. All user similarities are computed from the observed interactions in the train set (i.e., the binarised interaction matrix). We compute PUF with P and NDCG, to represent set- and rank-based measures, as well as for consistency with other \textsc{Fair} measures which are also computed with P and NDCG. To avoid extremely small values, we min-max normalise the pairwise user similarity per dataset.

\subsection{Comparison of all evaluation measures}
\label{ss:groundwork}
\Cref{tab:base} shows the \textsc{Eff} and \textsc{Fair} evaluation results. To identify empirical limitations in the measures, we study their score range and computational efficiency.

\begin{table}[t]
\caption{Statistics of the preprocessed datasets.}
\centering
\resizebox{0.7\columnwidth}{!}{
\begin{tabular}{lrrrr}
\toprule
\textbf{dataset} & \multicolumn{1}{l}{\textbf{\#users (all/test)}} & \multicolumn{1}{l}{\textbf{\#items}} & \multicolumn{1}{l}{\textbf{\#interactions}} & \multicolumn{1}{l}{\textbf{sparsity (\%)}} \\ 
\midrule                                                                    
Lastfm \cite{Cantador20112nd2011}& 1,859/1,836 & 2,823 & 71,355 & 98.64\% \\
QK-video \cite{Yuan2022Tenrec:Systems} & 4,656/3,514 & 6,423 & 51,777 & 99.83\% \\
ML-10M \cite{Harper2015TheContext} & 49,378/1,523 & 9,821 & 5,362,685 & 98.89\% \\
ML-20M \cite{Harper2015TheContext} & 89,917/2,178 & 16,404 & 10,588,141 & 99.28\% \\
\bottomrule
\end{tabular}
}
\label{tab:stats}
\end{table}
\begin{table}[t]
\centering
\caption{\textsc{Eff} and \textsc{Fair} scores at $k=10$ for recommenders. The most effective/fair score per measure is bolded. $\uparrow$/$\downarrow$ means the higher/lower the better.}
\label{tab:base}
\begin{minipage}[t]{.495\textwidth}
\resizebox{0.99\columnwidth}{!}{
\begin{tabular}{lllrrrrrrr}
\toprule
 &  &  & U-KNN & I-KNN & BPR & NMF & MVAE & NGCF & NCL \\
\midrule
\multirow[c]{18}{*}{\rotatebox[origin=c]{90}{Lastfm}} & \multirow[c]{6}{*}{\rotatebox[origin=c]{90}{\textsc{Eff}}} & $\uparrow$ $\text{HR}$ & 0.7102 & 0.7647 & 0.7729 & 0.7424 & 0.7783 & 0.7707 & \bfseries 0.7930 \\
 &  & $\uparrow$ $\text{MRR}$ & 0.4386 & 0.4835 & 0.4916 & 0.4641 & 0.4764 & 0.4954 & \bfseries 0.5026 \\
 &  & $\uparrow$ $\text{P}$ & 0.1532 & 0.1721 & 0.1782 & 0.1649 & 0.1758 & 0.1779 & \bfseries 0.1840 \\
 &  & $\uparrow$ $\text{MAP}$ & 0.1190 & 0.1369 & 0.1414 & 0.1291 & 0.1377 & 0.1418 & \bfseries 0.1478 \\
 &  & $\uparrow$ $\text{R}$ & 0.1899 & 0.2180 & 0.2238 & 0.2069 & 0.2241 & 0.2236 & \bfseries 0.2342 \\
 &  & $\uparrow$ $\text{NDCG}$ & 0.2167 & 0.2450 & 0.2520 & 0.2334 & 0.2473 & 0.2530 & \bfseries 0.2611 \\
\cline{2-10}
 & \multirow[c]{8}{*}{\rotatebox[origin=c]{90}{\textsc{Fair} (existing)}} & $\downarrow$ $\text{SD-P}$ & \bfseries 0.1485 & 0.1529 & 0.1538 & 0.1505 & 0.1518 & 0.1527 & 0.1552 \\
 &  & $\downarrow$ $\text{SD-NDCG}$ & \bfseries 0.2040 & 0.2091 & 0.2092 & 0.2072 & 0.2084 & 0.2085 & 0.2124 \\
 &  & $\downarrow$ $\text{Gini-P}$ & 0.5146 & 0.4768 & 0.4668 & 0.4889 & 0.4654 & 0.4642 & \bfseries 0.4554 \\
 &  & $\downarrow$ $\text{Gini-NDCG}$ & 0.5213 & 0.4789 & 0.4678 & 0.4953 & 0.4735 & 0.4646 & \bfseries 0.4581 \\
 & & $\downarrow$ $\text{ME}$ & 0.0897 & \bfseries 0.0827 & 0.0904 & 0.0983 & 0.0973 & 0.0897 & 0.0886 \\
 &  & $\downarrow$ $\text{MME}$ & 0.0025 & \bfseries 0.0017 & 0.0020 & 0.0025 & 0.0020 & 0.0021 & 0.0019 \\
 &  & $\downarrow$ $\text{PEU}$ & 0.6525 & \bfseries 0.6176 & 0.6318 & 0.6765 & 0.6531 & 0.6275 & 0.6193 \\
 & & $\downarrow$ $\text{UF}$ & \bfseries 0.6615 & 0.6616 & 0.6660 & 0.6667 & 0.6685 & 0.6675 & 0.6681 \\
\cline{2-10}
 & \multirow[c]{4}{*}{\rotatebox[origin=c]{90}{\scriptsize\textsc{Fair} (PUF)}} &  $\downarrow$ $\text{Prec-Cos}$ & 0.0097 & 0.0100 & 0.0100 & \bfseries 0.0096 & 0.0098 & 0.0099 & 0.0100 \\
 &  & $\downarrow$ $\text{Prec-Jacc}$ & 0.0073 & 0.0075 & 0.0075 & \bfseries 0.0073 & 0.0074 & 0.0075 & 0.0076 \\
 &  & $\downarrow$ $\text{NDCG-Cos}$ & 0.0134 & 0.0137 & 0.0136 & \bfseries 0.0132 & 0.0135 & 0.0135 & 0.0136 \\
 &  & $\downarrow$ $\text{NDCG-Jacc}$ & 0.0101 & 0.0103 & 0.0102 & \bfseries 0.0099 & 0.0101 & 0.0101 & 0.0102 \\
\cline{1-10} 
\multirow[c]{18}{*}{\rotatebox[origin=c]{90}{QK-video}} & \multirow[c]{6}{*}{\rotatebox[origin=c]{90}{\textsc{Eff}}} & $\uparrow$ $\text{HR}$ & 0.1155 & 0.0396 & 0.0993 & 0.0857 & 0.1093 & 0.1238 & \bfseries 0.1298 \\
 &  & $\uparrow$ $\text{MRR}$ & 0.0435 & 0.0131 & 0.0390 & 0.0303 & 0.0392 & \bfseries 0.0484 & 0.0481 \\
 &  & $\uparrow$ $\text{P}$ & 0.0122 & 0.0041 & 0.0105 & 0.0088 & 0.0116 & 0.0134 & \bfseries 0.0140 \\
 &  & $\uparrow$ $\text{MAP}$ & 0.0188 & 0.0046 & 0.0170 & 0.0128 & 0.0178 & \bfseries 0.0218 & 0.0218 \\
 &  & $\uparrow$ $\text{R}$ & 0.0517 & 0.0141 & 0.0433 & 0.0369 & 0.0506 & 0.0589 & \bfseries 0.0608 \\
 &  & $\uparrow$ $\text{NDCG}$ & 0.0331 & 0.0091 & 0.0290 & 0.0231 & 0.0314 & 0.0375 & \bfseries 0.0381 \\
\cline{2-10}
 & \multirow[c]{8}{*}{\rotatebox[origin=c]{90}{\textsc{Fair} (existing)}} & $\downarrow$ $\text{SD-P}$ & 0.0350 & \bfseries 0.0205 & 0.0327 & 0.0293 & 0.0343 & 0.0373 & 0.0376 \\
 &  & $\downarrow$ $\text{SD-NDCG}$ & 0.1089 & \bfseries 0.0539 & 0.1058 & 0.0910 & 0.1079 & 0.1191 & 0.1188 \\
 &  & $\downarrow$ $\text{Gini-P}$ & 0.8906 & 0.9618 & 0.9060 & 0.9168 & 0.8968 & 0.8850 & \bfseries 0.8790 \\
 &  & $\downarrow$ $\text{Gini-NDCG}$ & 0.9215 & 0.9733 & 0.9344 & 0.9425 & 0.9265 & 0.9156 & \bfseries 0.9120 \\
 & & $\downarrow$ $\text{ME}$ & 0.0874 & 0.0957 & 0.0962 & 0.0757 & \bfseries 0.0412 & 0.0899 & 0.0622 \\
 &  & $\downarrow$ $\text{MME}$ & 0.0012 & \bfseries 0.0003 & 0.0010 & 0.0014 & 0.0011 & 0.0011 & 0.0012 \\
 &  & $\downarrow$ $\text{PEU}$ & 0.7746 & 0.9087 & 0.8426 & 0.6810 & \bfseries 0.3867 & 0.8056 & 0.5626 \\
 & & $\downarrow$ $\text{UF}$ & 0.5731 & 0.5751 & 0.5739 & 0.5736 & \bfseries 0.5721 & 0.5734 & 0.5726 \\
\cline{2-10}
 & \multirow[c]{4}{*}{\rotatebox[origin=c]{90}{\scriptsize\textsc{Fair} (PUF)}} &  $\downarrow$ $\text{Prec-Cos}$ & 0.0001 & \bfseries 0.0000 & 0.0001 & 0.0001 & 0.0001 & 0.0001 & 0.0001 \\
 &  & $\downarrow$ $\text{Prec-Jacc}$ & 0.0001 & \bfseries 0.0000 & 0.0001 & 0.0001 & 0.0001 & 0.0001 & 0.0001 \\
 &  & $\downarrow$ $\text{NDCG-Cos}$ & 0.0003 & \bfseries 0.0001 & 0.0003 & 0.0002 & 0.0003 & 0.0003 & 0.0003 \\
 &  & $\downarrow$ $\text{NDCG-Jacc}$ & 0.0002 & \bfseries 0.0001 & 0.0002 & 0.0002 & 0.0002 & 0.0002 & 0.0002 \\
\bottomrule
\end{tabular}
}
\end{minipage}%
\noindent\begin{minipage}[t]{.495\textwidth}
\resizebox{0.985\linewidth}{!}{
\begin{tabular}{lllrrrrrrr}
\toprule
 &  &  & U-KNN & I-KNN & BPR & NMF & MVAE & NGCF & NCL \\
\cline{1-10}
\multirow[c]{18}{*}{\rotatebox[origin=c]{90}{ML-10M}} & \multirow[c]{6}{*}{\rotatebox[origin=c]{90}{\textsc{Eff}}} & $\uparrow$ $\text{HR}$ & 0.5207 & 0.4872 & 0.5121 & 0.5141 & 0.4169 & 0.5128 & \bfseries 0.5213 \\
 &  & $\uparrow$ $\text{MRR}$ & \bfseries 0.3105 & 0.2818 & 0.2987 & 0.2915 & 0.2372 & 0.2865 & 0.3019 \\
 &  & $\uparrow$ $\text{P}$ & 0.1531 & 0.1369 & 0.1458 & 0.1433 & 0.1066 & 0.1460 & \bfseries 0.1536 \\
 &  & $\uparrow$ $\text{MAP}$ & \bfseries 0.1027 & 0.0887 & 0.0953 & 0.0911 & 0.0666 & 0.0930 & 0.1009 \\
 &  & $\uparrow$ $\text{R}$ & \bfseries 0.0276 & 0.0222 & 0.0249 & 0.0246 & 0.0203 & 0.0246 & 0.0263 \\
 &  & $\uparrow$ $\text{NDCG}$ & \bfseries 0.1691 & 0.1501 & 0.1600 & 0.1560 & 0.1191 & 0.1579 & 0.1673 \\
\cline{2-10}
 & \multirow[c]{8}{*}{\rotatebox[origin=c]{90}{\textsc{Fair} (existing)}} & $\downarrow$ $\text{SD-P}$ & 0.2206 & 0.2080 & 0.2125 & 0.2076 & \bfseries 0.1781 & 0.2109 & 0.2195 \\
 &  & $\downarrow$ $\text{SD-NDCG}$ & 0.2376 & 0.2244 & 0.2297 & 0.2241 & \bfseries 0.1960 & 0.2262 & 0.2367 \\
 &  & $\downarrow$ $\text{Gini-P}$ & 0.6884 & 0.7076 & 0.6906 & 0.6874 & 0.7415 & 0.6880 & \bfseries 0.6853 \\
 &  & $\downarrow$ $\text{Gini-NDCG}$ & \bfseries 0.6889 & 0.7123 & 0.6954 & 0.6948 & 0.7460 & 0.6943 & 0.6911 \\
 & & $\downarrow$ $\text{ME}$ & 0.1017 & 0.1131 & 0.1290 & 0.1307 & \bfseries 0.0726 & 0.1284 & 0.1192 \\
 &  & $\downarrow$ $\text{MME}$ & 0.0039 & 0.0040 & 0.0046 & 0.0046 & \bfseries 0.0025 & 0.0043 & 0.0044 \\
 &  & $\downarrow$ $\text{PEU}$ & 0.5936 & 0.6356 & 0.6927 & 0.7072 & \bfseries 0.4668 & 0.7026 & 0.6619 \\
 & & $\downarrow$ $\text{UF}$ & \bfseries 0.6948 & 0.6954 & 0.7148 & 0.7103 & 0.6994 & 0.7095 & 0.7053 \\
\cline{2-10}
 & \multirow[c]{4}{*}{\rotatebox[origin=c]{90}{\scriptsize\textsc{Fair} (PUF)}} &  $\downarrow$ $\text{Prec-Cos}$ & 0.0493 & 0.0445 & 0.0472 & 0.0458 & \bfseries 0.0363 & 0.0469 & 0.0497 \\
 &  & $\downarrow$ $\text{Prec-Jacc}$ & 0.0298 & 0.0268 & 0.0285 & 0.0276 & \bfseries 0.0217 & 0.0284 & 0.0302 \\

 &  & $\downarrow$ $\text{NDCG-Cos}$ & 0.0547 & 0.0495 & 0.0523 & 0.0507 & \bfseries 0.0409 & 0.0515 & 0.0549 \\
 &  & $\downarrow$ $\text{NDCG-Jacc}$ & 0.0331 & 0.0298 & 0.0316 & 0.0306 & \bfseries 0.0245 & 0.0311 & 0.0333 \\
\cline{1-10} 
\multirow[c]{18}{*}{\rotatebox[origin=c]{90}{ML-20M}} & \multirow[c]{6}{*}{\rotatebox[origin=c]{90}{\textsc{Eff}}} & $\uparrow$ $\text{HR}$ & \bfseries 0.5092 & 0.4881 & 0.5051 & 0.4927 & 0.4890 & 0.5083 & 0.5051 \\
 &  & $\uparrow$ $\text{MRR}$ & 0.2970 & 0.2799 & 0.2928 & 0.2781 & 0.2593 & \bfseries 0.3007 & 0.2934 \\
 &  & $\uparrow$ $\text{P}$ & \bfseries 0.1582 & 0.1423 & 0.1449 & 0.1433 & 0.1409 & 0.1517 & 0.1500 \\
 &  & $\uparrow$ $\text{MAP}$ & \bfseries 0.1079 & 0.0923 & 0.0957 & 0.0934 & 0.0889 & 0.1010 & 0.0997 \\
 &  & $\uparrow$ $\text{R}$ & \bfseries 0.0213 & 0.0193 & 0.0193 & 0.0186 & 0.0185 & 0.0199 & 0.0200 \\
 &  & $\uparrow$ $\text{NDCG}$ & \bfseries 0.1711 & 0.1538 & 0.1584 & 0.1544 & 0.1481 & 0.1649 & 0.1626 \\
\cline{2-10}
 & \multirow[c]{8}{*}{\rotatebox[origin=c]{90}{\textsc{Fair} (existing)}} & $\downarrow$ $\text{SD-P}$ & 0.2327 & \bfseries 0.2145 & 0.2157 & 0.2160 & 0.2152 & 0.2239 & 0.2228 \\
 &  & $\downarrow$ $\text{SD-NDCG}$ & 0.2484 & 0.2290 & 0.2333 & 0.2324 & \bfseries 0.2278 & 0.2399 & 0.2385 \\
 &  & $\downarrow$ $\text{Gini-P}$ & \bfseries 0.6981 & 0.7068 & 0.6991 & 0.7054 & 0.7100 & 0.6983 & 0.7001 \\
 &  & $\downarrow$ $\text{Gini-NDCG}$ & \bfseries 0.7021 & 0.7110 & 0.7053 & 0.7138 & 0.7203 & 0.7022 & 0.7046 \\
 & & $\downarrow$ $\text{ME}$ & \bfseries 0.1034 & 0.1199 & 0.1361 & 0.1412 & 0.1421 & 0.1348 & 0.1337 \\
 &  & $\downarrow$ $\text{MME}$ & \bfseries 0.0038 & 0.0049 & 0.0046 & 0.0052 & 0.0049 & 0.0046 & 0.0050 \\
 &  & $\downarrow$ $\text{PEU}$ & \bfseries 0.5877 & 0.6327 & 0.7029 & 0.6956 & 0.7071 & 0.6910 & 0.6869 \\
 & & $\downarrow$ $\text{UF}$ & 0.7098 & \bfseries 0.7098 & 0.7339 & 0.7305 & 0.7336 & 0.7309 & 0.7246 \\
\cline{2-10}
 & \multirow[c]{4}{*}{\rotatebox[origin=c]{90}{\scriptsize\textsc{Fair} (PUF)}} & $\downarrow$ $\text{Prec-Cos}$ & 0.0553 & 0.0502 & 0.0504 & 0.0503 & \bfseries 0.0497 & 0.0529 & 0.0525 \\
 &  & $\downarrow$ $\text{Prec-Jacc}$ & 0.0329 & 0.0298 & 0.0299 & 0.0299 & \bfseries 0.0296 & 0.0314 & 0.0312 \\
 &  & $\downarrow$ $\text{NDCG-Cos}$ & 0.0605 & 0.0548 & 0.0559 & 0.0552 & \bfseries 0.0532 & 0.0581 & 0.0576 \\
 &  & $\downarrow$ $\text{NDCG-Jacc}$ & 0.0361 & 0.0326 & 0.0333 & 0.0328 & \bfseries 0.0317 & 0.0346 & 0.0343 \\
\bottomrule
\end{tabular}
}
\end{minipage}
\end{table}
\begin{table}[t]
\centering 
\caption{Mean computation time (s) of \textsc{Fair} measures per model.}
\label{tab:time}
\resizebox{\columnwidth}{!}{
\begin{tabular}{lrrrrrrrr|rrrr}
\toprule
& \multicolumn{8}{c|}{Existing measures} & \multicolumn{4}{c}{PUF}\\
 \midrule
 & SD-P & SD-NDCG & Gini-P & Gini-NDCG & ME & MME & PEU & UF & Prec-Cos & Prec-Jacc & NDCG-Cos & NDCG-Jacc \\
\midrule
Lastfm & <0.01 & <0.01 & <0.01 & <0.01 & 233.88 & 233.91 & 233.91 & 1297.78 & 5.30 & 9.79 & 4.32 & 9.56 \\
QK-video & 0.01 & 0.02 & 0.01 & 0.03 & 1867.81 & 1867.88 & 1867.88 & 873.52 & 20.68 & 35.63 & 16.01 & 35.10 \\
ML-10M & <0.01 & 0.01 & <0.01 & 0.02 & 871.17 & 871.20 & 871.20 & 1020.16 & 4.00 & 7.22 & 3.38 & 7.40 \\
ML-20M & <0.01 & <0.01 & <0.01 & <0.01 & 2020.60 & 2020.70 & 2020.70 & 1358.52 & 8.25 & 15.65 & 6.72 & 15.19 \\
\bottomrule
\end{tabular}}
\end{table}

\noindent \textbf{Score range}. A measure that does not have a wide score range across various fairness levels is less useful in distinguishing changes in fairness. Such measures may create an illusion of a negligible difference in fairness, due to their compressed empirical range \cite{Rampisela2024CanRelevance,Rampisela2025RelevanceGuidelines,10.1145/3459637.3482287,10.1145/3121050.3121072}. Across datasets, the observed range (not the theoretical range) of each \textsc{Fair} measure varies, except for MME, which is extremely small ($\leq5.2\times10^{-3}$). Swapping a user's recommendation list with another user's does not generally result in a large increase in the user's P@$k$ (envy), which translates to low MME. ME and PEU are unaffected by this despite being envy-based, as ME accounts for envy across all users rather than the maximum envy per user, and PEU employs an envy threshold. 
In short, MME is the least sensitive as it fails to discriminate between models, while other \textsc{Fair} measures, including PUF, are more sensitive. The most sensitive measures, i.e., the ones with the widest observed range, are Gini and PEU.

\noindent \textbf{Measure computation time}. We report the average computation time of the \textsc{Fair} measures in \Cref{tab:time}. All runs are done with AMD EPYC 75F3 for a fair comparison. User pairwise similarity is computed once per dataset for each PUF variant. We find that half of the existing measures, i.e., ME, MME, PEU, and UF are computationally expensive ($>4$ mins.), while PUF is significantly faster ($<40$ s), despite also being a pairwise measure. ME/MME/PEU/UF need additional extensive computation per pair, which makes them expensive: UF does nested pairwise comparisons, while ME/MME/PEU recompute the effectiveness score for each user pair. In short, PUF is computationally more efficient and thus more practical to compute than most existing \textsc{Fair} measures.

\subsection{Measure agreement}
\label{ss:agreement}

\begin{figure}[tb]
  \includegraphics[width=\textwidth, trim=0.25cm 0.25cm 0.25cm 0.1cm, clip=True]{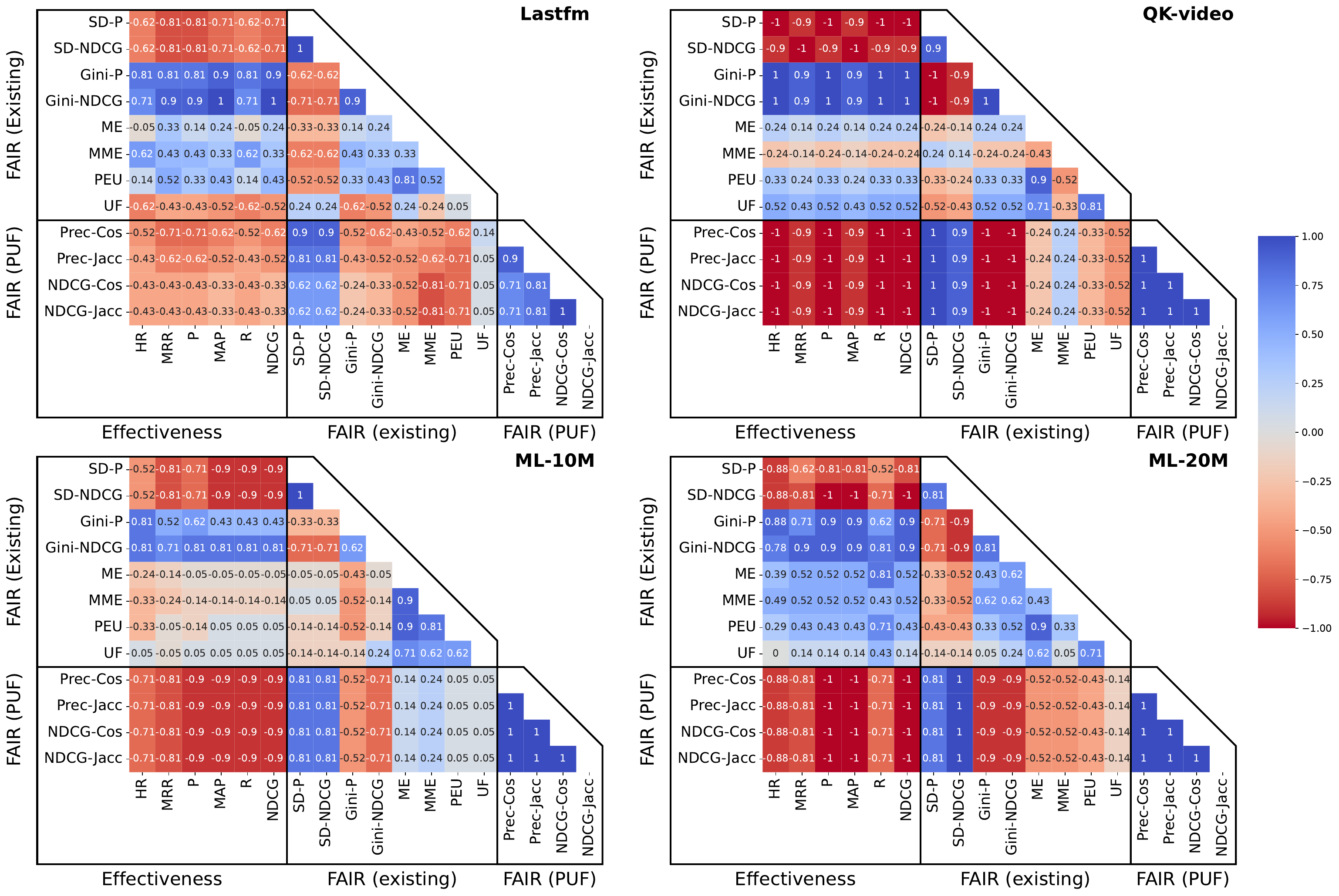}  
    \caption{Kendall's $\tau$ correlation between all measures (\textsc{Eff}, \textsc{Fair}, and PUFs).}
    \label{fig:corr}
\end{figure}

An important aspect when comparing evaluation measures is how much they agree when their scores are used to rank models from best to worst. 
If one measure can be used to estimate the rank ordering given by another, there is no point in using both measures if we are only interested in ranking models. To study this, we compute Kendall's $\tau$ correlation for all \textsc{Eff} and \textsc{Fair} measures. Kendall's $\tau$ can handle ties (unlike Spearman's $\rho$) and is more robust to nonlinear relationships (unlike Pearson's coefficient). 
If two measures have $\tau\geq 0.9$, we consider their rankings equivalent \cite{Voorhees2001EvaluationDocuments}. \Cref{fig:corr} shows the agreement between (i) \textsc{Eff} and \textsc{Fair} measures, and (ii) among \textsc{Fair} measures.  
 
\noindent \textbf{Agreement between \textsc{Eff} and \textsc{Fair} measures}. Across datasets, the agreement between \textsc{Fair} and \textsc{Eff} measures varies from strong disagreement (e.g., SD with $\tau \in [-1, -0.52]$) to moderate-to-strong agreement (e.g., Gini with $\tau \in [0.43, 1]$). Our PUF consistently disagrees with \textsc{Eff} measures ($\tau\leq-0.71$), even if the disagreement is weaker for Lastfm, $\tau \in [-0.71,-0.33]$. As no \textsc{Fair} measure consistently has $|\tau| \geq 0.9$ with any \textsc{Eff} measure for all datasets, their model orderings cannot be precisely inferred from that of \textsc{Eff} measures.

\noindent \textbf{Agreement among \textsc{Fair} measures}. 
Regarding the agreement between PUF and existing \textsc{Fair} measures, we find that only SD aligns with PUF ($\tau \geq 0.62$), while the rest tend to disagree or have a weak correlation with PUF, $\tau \in [-1, 0.24]$. Both UF and PUF consider user similarity, yet UF weakly correlates to PUF ($\tau \in [-0.14, 0.14]$) for all datasets except QK-video ($\tau =-0.52$). This weak relationship may be due to UF not considering item relevance, while PUF does. Even if we only compare the best model based on the measures instead of the model rankings, UF and PUF never agree. Thus, using UF and PUF to gauge fairness may differ in the conclusion on which model is the fairest. 
In contrast, despite not being similarity-based, SD correlates the strongest to PUF. Yet, most of its rankings are not equivalent to PUF ($\tau < 0.9$), which means that fairness evaluation with SD can lead to a conclusion that misaligns with PUF.

Overall, \textsc{Fair} measures often disagree on model orderings with PUF, regardless of whether the measure accounts for user similarity. While SD has the most similar conclusions to PUF, it still does not give equivalent rankings to PUF. 

\subsection{Varying relevance score distribution}
\label{ss:artificial-rel}
\begin{figure*}[t]
    \includegraphics[width=\textwidth, trim=0.25cm 0.2cm 0.25cm 0.2cm, clip=True]{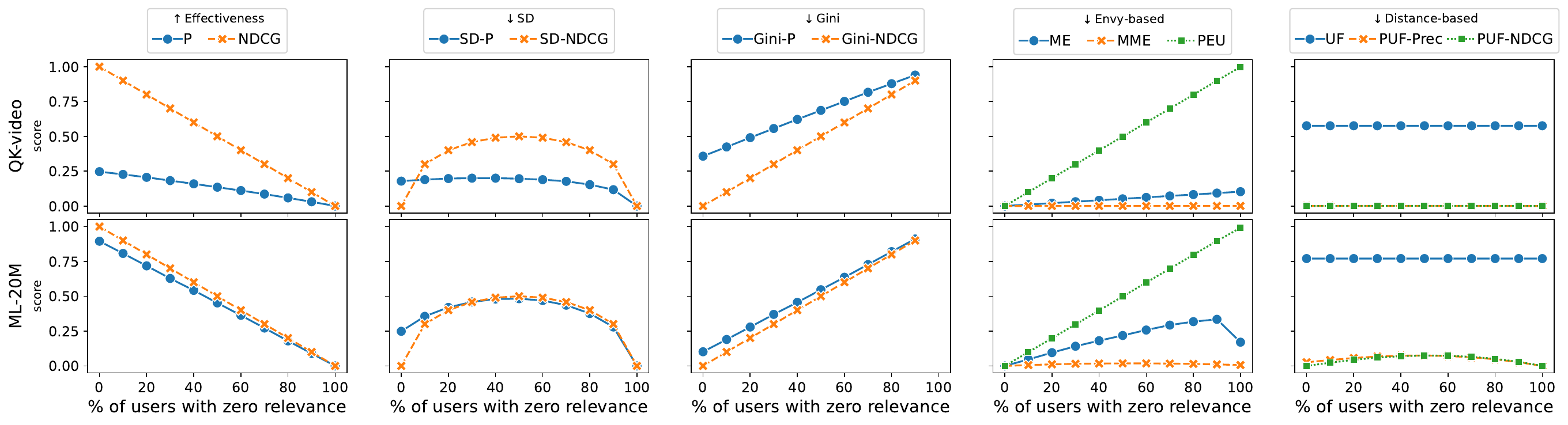}
    \caption{
    Effectiveness (\textsc{Eff}) and fairness (\textsc{Fair}) scores of QK-video and ML-20M, when artificially varying \% of users with all irrelevant items (zero relevance), and the rest of the users receiving all relevant items. All PUF variants overlap. Gini is missing points at 100\% users with zero relevance as it is undefined when each user has zero \textsc{Eff} scores. 
    }
    \label{fig:artificial-relevance}
\end{figure*}
\begin{figure}[t]
\centering
\begin{minipage}[t]{.48\textwidth}
 \includegraphics[width=\columnwidth, 
        trim=0.3cm 0.2cm 0.2cm 0.2cm, clip=True]{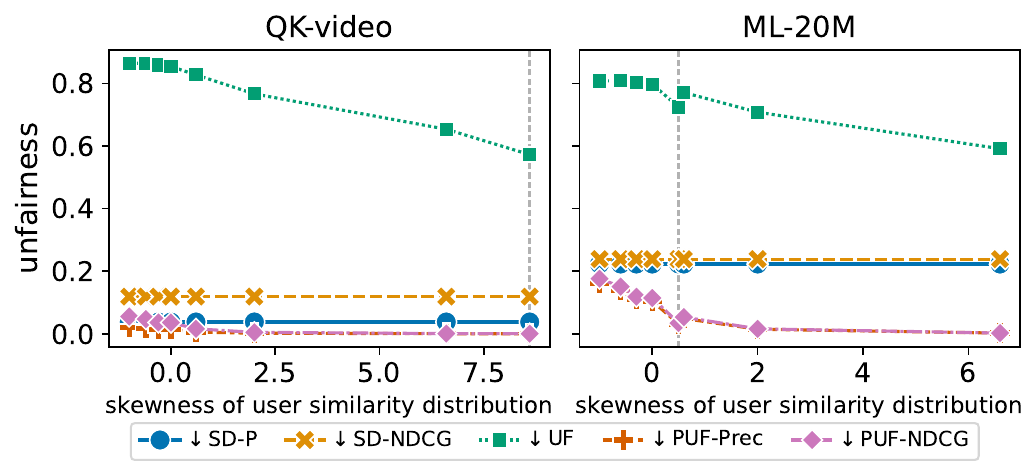}
    \caption{Artificially varying the skewness of the user similarity distribution for QK-video and ML-20M. Vertical grey lines denote the skewness corresponding to $sim_{Jacc}$ observed in the dataset. The distribution skewness differs across datasets.
    }
    \label{fig:artificial-similarity}   
\end{minipage}\hspace{5pt}
\begin{minipage}[t]{.48\textwidth}
\centering
    \includegraphics[width=\linewidth, trim=0.2cm 0.2cm 0.2cm 0.2cm, clip=True]{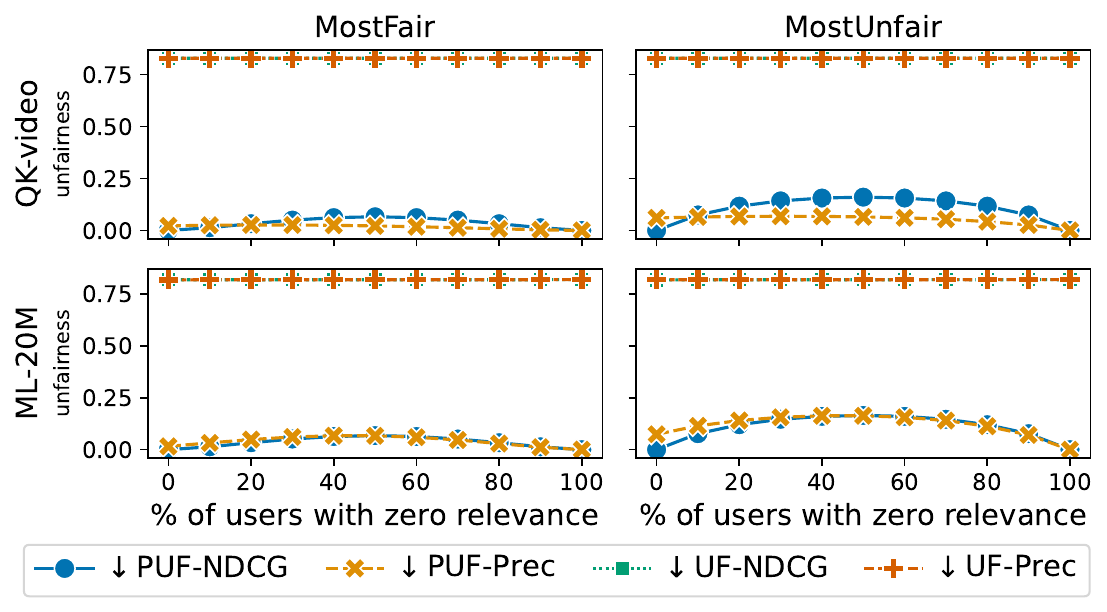}
    \caption{Artificially varying the \% of users with zero relevance for QK-video and ML-20M. Lower \textsc{Eff} score difference is assigned to user pairs with higher similarity (MostFair), and to lower similarity (MostUnfair). Both UF overlap.}  
    \label{fig:artificial-sim-relevance-sort}

\end{minipage}
\end{figure}

It is important that a fairness measure captures differences in recommendation quality across users, as fairness is related to the disparity in recommendation (relevance). So, any change in \textsc{Eff} scores should also be reflected in the \textsc{Fair} scores. We thus study how varying item relevance affects \textsc{Eff} and \textsc{Fair} scores. All \textsc{Fair} measure equations, except UF, explicitly account for effectiveness, but it is unknown how sensitive \textsc{Fair} measures are to changes in \textsc{Eff} scores. 

\noindent\textbf{Procedure.} The change of relevance scores is artificially done as follows. 
For all users, we start by recommending $k$ relevant items (based on the test set). For users with more than $k$ relevant items, we select the relevant items randomly. For users with fewer than $k$ relevant items, we fill the remaining slots with random irrelevant items to ensure that each user receives exactly $k$ items. In each iteration, we replace the recommendation of 10\% of the users with all irrelevant items and compute the measures. We expect maximum fairness at the start (as all users have the maximum \textsc{Eff} scores\footnote{P@$k$-based scores may not be optimal at the start as some users have $<k$ test items.}) and at the end (as all users have 0 \textsc{Eff} scores). We expect maximum unfairness when half the users get all relevant items, and the rest get irrelevant items, as this leads to one of the most uneven \textsc{Eff} score distributions.
We only compute PUF with $sim_{Jacc}$, as the model orderings given by PUF-*-Cos are equivalent to PUF-*-Jacc (\Cref{ss:agreement}). User similarities are computed based on the observed interactions in the train sets.
 
\noindent\textbf{Results.} \Cref{fig:artificial-relevance} shows the results for QK-video and ML-20M, which represent the overall trends in all our datasets (the rest are in the online appendix). As expected, PUF has decreasing fairness followed by increasing fairness, shown by the inverted parabolas. This trend is more pronounced for the ML-* datasets, as the mean user pairwise similarity is higher than for the other datasets. Among the \textsc{Fair} measures, only SD follows this expectation; the others show undesirable tendencies: as \textsc{Eff} drops, Gini and PEU notably become less fair, and ME also but to a minor extent. The overall fairness drop in these measures is undesirable, as the scores closely resemble decreasing effectiveness instead of the disparity in the \textsc{Eff} scores distribution. 
Even worse, MME and UF are almost invariant to the change in recommendation effectiveness: MME tends to score extremely low to begin with, while UF does not depend on item relevance.

To sum up, PUF and SD quantify fairness based on the disparity in recommendation effectiveness. All other \textsc{Fair} measures ignore disparity; they just reflect effectiveness drops or are insensitive to changes in effectiveness.
Next, we ask if changes in user similarity distribution are reflected in the \textsc{Fair} measures.

\subsection{Varying user similarity distribution}
\label{ss:artificial-sim}

Individual user fairness is defined based on user similarity; it is important to know how user similarity affects fairness. While two/more recommender models should be evaluated under the same similarity distribution, a desirable individual user fairness measure should be able to distinguish a single model's performance across different similarity distributions, which may arise from various ways of modelling user similarity. To this end, we investigate how the \textsc{Fair} measures respond to (artificial) variations of user similarity distributions.

\noindent\textbf{Procedure.} User similarity distributions tend to be right-skewed (many dissimilar users) for random users, and left-skewed (many similar users) for users that are friends  \cite{Reisz2024QuantifyingPrediction}. Further, users are often dissimilar, as some users are new to the systems or do not engage much, leading to discrepancies in the number of interactions among users and potentially affecting user similarity. Considering the above, we create synthetic user similarity scores by sampling from the Weibull distribution \cite{Weibull1951AApplicability}, which can be used to model skewed distributions. It has been used to model user rating distributions and sampling user neighbour candidates in RSs \cite{Kermany2020ReInCre:Credibility,Kermany2023IncorporatingSystems,Adamopoulos2014OnSystems}. Its probability density function is $p(x) = \lambda x^{\lambda-1} \exp{(-x^\lambda)}$. To obtain various right- and left-skewed distributions to represent possible user similarity distributions, we set $\lambda \in \{0.5, 1, 2, 5, 10, 50\}$. We also sample from the normal distribution $\mathcal{N}(0,1)$, which has zero skews (i.e., equal portions of user pairs with similarity below/above the mean). We then min-max normalise the sampled similarity values to rescale them in the $[0,1]$-range and randomly assign them to user pairs. We analyse non-random assignment in \Cref{ss:assignment}.

While the user similarity is artificial, we use the actual recommendation lists and scores from the NCL model runs as they perform relatively well. To save computation time, we only compute SD to represent all similarity-independent measures; theoretically, these measures will remain constant given no change in their input. We also compute PUF and UF, the two similarity-based measures. We compare the measure scores with the scores corresponding to the user similarity distribution observed in the datasets based on $sim_{Jacc}$ (\Cref{ss:groundwork}).

\noindent\textbf{Results.} \Cref{fig:artificial-similarity} shows the results for QK-video and ML-20M (the rest have similar trends and are in the online appendix). We see that the similarity-based measures PUF and UF become fairer as skewness increases. Increasing skewness means a higher proportion of user pairs with low similarity, hence $\downarrow$PUF and $\downarrow$UF tend to be lower. Conversely, but as expected, the similarity-independent SD remains constant despite the change in skewness. For highly skewed similarities (skewness $>6$), which is a realistic similarity distribution as seen in QK-video, SD-NDCG is somewhat unfair ($\approx0.2$) for all datasets. Yet, PUF is almost perfectly fair ($\approx0$). Therefore, simply using SD may lead to the underestimation of fairness. While we only compute SD here, we expect the other \textsc{Fair} measures to show the same invariance as they are similarity-independent. 

Between the two similarity-based measures, PUF is more sensitive to negatively skewed similarity distribution than UF. As skewness decreases, the mean user similarity increases at a slower rate. UF only considers user pairs above the mean-based similarity threshold, thus the number of user pairs contributing to its log sum decreases slower than in the right-skewed distributions. Another concern of UF is its relatively high unfairness compared to PUF and SD, even when most users are dissimilar. This may be because UF computes the pairwise distance of the representation of the recommended items of a user pair. Minimising this distance is hard as each recommended item of a user must have a similar representation to each of the other user's items, regardless of item relevance.

To sum up, PUF can distinguish fairness levels across various similarity distributions, while non-similarity-based measures cannot. 
This shows the strength of PUF over SD (and indirectly over other non-similarity-based measures, i.e., all \textsc{Fair} measures except UF). We find that disregarding user similarity can also lead to the misinterpretation of fairness level. Next, we compare the two similarity-based measures and show the strengths of PUF over UF.

\subsection{PUF and UF under extreme cases}
\label{ss:assignment}

We compare the similarity-based measures (PUF and UF) under extreme scenarios: can their scores reflect the difference in maximum and minimum fairness, across differences in the recommendation quality? Given an artificial set of pairwise user similarities and an artificial set of pairwise \textsc{Eff} score differences, to simulate the fairest case (MostFair), we sort these values and assign a higher similarity value to user pairs with lower \textsc{Eff} score difference. Separately, we assign a higher similarity score to pairs with higher \textsc{Eff} score difference to mimic the unfairest case (MostUnfair). For MostFair, a desirable measure would score close to 0 (the fairest). For MostUnfair, it should exhibit an inverted U-shape when the \textsc{Eff} score distribution is varied (i.e., similar to \Cref{fig:artificial-relevance}), as the maximum unfairness happens when the \textsc{Eff} distribution is the most uneven.

\noindent\textbf{Procedure.} We use an artificial, right-skewed user similarity distribution sampled from the Weibull distribution with $\lambda=2$ (\Cref{ss:artificial-sim}). We use the P@$k$ and NDCG@$k$ scores per user from the artificial runs in \Cref{ss:artificial-rel}. To compute PUF- and UF-NDCG, we assign the user similarities to user pairs following the sorted pairwise difference of NDCG, and likewise for PUF- and UF-Prec.

\noindent\textbf{Results.} \Cref{fig:artificial-sim-relevance-sort} shows the results for QK-video and ML-20M (the rest have similar trends and are in the online appendix). The MostFair assignment yields PUF scores that are overall close to the fairest (0) across varying recommendation effectiveness, but UF remains constantly unfair ($\approx$0.8) nonetheless. This emphasises a mismatch between fairness computed based on the disparity of item representation and based on \textsc{Eff} score differences. 
PUF scores are more unfair for the MostUnfair than the MostFair scenario, while UF is almost invariant to the change in similarity assignment between MostFair and MostUnfair. This confirms the insensitivity of UF towards varying recommendation effectiveness seen in \Cref{fig:artificial-relevance}. In the MostUnfair case, $\downarrow$PUF ranges in $[0,0.2)$ (close to the fairest) for all datasets. This is because the overall user similarity is low, which can happen in real-world scenarios. Hence, the maximum unfairness is expected to be relatively low, as PUF relies on user similarity to quantify fairness.

To sum up, PUF correctly scores close to the fairest for MostFair, while UF does not. PUF is notably more unfair for the MostUnfair than the MostFair case, while UF is almost constant for both cases and across varying effectiveness. In both cases, UF overestimates effectiveness-based unfairness by constantly scoring much higher than PUF. Overall, PUF can reliably measure extreme (un)fairness.

\section{Related work}
In \Cref{s:individual-user-fair} we overviewed existing individual user fairness measures. 
Here, we discuss work that studies RS fairness measures empirically \cite{Raj2022MeasuringResults,Rampisela2024EvaluationStudy,Rampisela2024CanRelevance,Rampisela2025RelevanceGuidelines} and proposes pairwise individual fairness measures for ranking \cite{Fabris2023PairwiseMeasure,Wang2022ProvidingSystems}. Our work is close to \cite{Raj2022MeasuringResults}, which studies item group fairness measures in RS, and \cite{Rampisela2024EvaluationStudy,Rampisela2024CanRelevance,Rampisela2025RelevanceGuidelines}, which studies evaluation measures of individual item fairness. Similarly to \cite{Raj2022MeasuringResults,Rampisela2024CanRelevance}, we find that fairness measures may disagree in their model ordering, and that some measures are more sensitive than others, given decreasing effectiveness and disparity in the recommendations. 
Among the individual fairness measures in \cite{Rampisela2024EvaluationStudy,Rampisela2024CanRelevance,Rampisela2025RelevanceGuidelines}, the similarity criterion of individual fairness is often ignored. Recent work \cite{Wang2022ProvidingSystems,Fabris2023PairwiseMeasure} also proposes pairwise individual fairness measures, but these are for items, whereas our pairwise measure, PUF, is for individual users. Further, PUF considers user similarity, while the measure in \cite{Fabris2023PairwiseMeasure} does not. The measure in \cite{Wang2022ProvidingSystems} is similar to UF, as both employ thresholding of user similarities. Instead of applying a threshold, which can be arbitrary, PUF is weighted by user similarities, which introduces degrees of user similarity in the fairness computation. There exists also work on user group fairness (e.g., \cite{Ekstrand2018AllEffectiveness,Zhu2018Fairness-AwareRecommendation}) or counterfactual fairness (e.g., \cite{Chen2024FairGap:Graph}). Most such work requires sensitive attributes (e.g., gender), but public recommendation datasets with sensitive attributes tend to lack user representations (e.g., only binary genders) \cite{Harper2015TheContext,Celma2010MusicTail,Yuan2022Tenrec:Systems}, and grouping users may require discretising the attribute (e.g., age) \cite{Buyl2024InherentFairness}.  
We focus on attribute-free individual fairness, rather than group or counterfactual fairness, to better assess distribution across all individuals \cite{Lazovich2022MeasuringMetrics} (often hidden in group fairness evaluation \cite{Fabris2023PairwiseMeasure,Rampisela2025StairwayFairness}). 

\section{Discussion and conclusions}
Current evaluation measures of individual user fairness in RSs consider either only disparity in recommendation effectiveness or user similarity, but never both jointly. None of them aligns with both the individual fairness definition and user utility as a key objective of RSs. To address this issue, we introduced PUF, a novel evaluation measure that quantifies user fairness through pairwise difference in effectiveness scores and considers similarity between users. While PUF is simple, it is a novel, intuitive measure that is robust against various effectiveness disparity or user similarity, which sets it apart from existing metrics, addressing a crucial gap in RS fairness evaluation. We recommend using PUF to evaluate individual user fairness due to its alignment with individual fairness definition, computational efficiency, and sensitivity to varying levels of user similarities and recommendation effectiveness in both typical and extreme cases. Future work includes integrating graded relevance or other ways of modelling user similarity. 

\begin{credits}
\subsubsection{\ackname}
The work is supported by the Algorithms, Data, and Democracy project (ADD-project), funded by the Villum Foundation and Velux Foundation.
We thank the anonymous reviewers who have provided helpful feedback to improve earlier versions of the manuscript.
\end{credits}

\bibliographystyle{splncs04}
\bibliography{references}

\end{document}